\documentclass[a4paper,pdftex,12pt]{article}

\usepackage{ifpdf}
\usepackage{geometry}
\usepackage[utf8]{inputenc} 
\usepackage[T1]{fontenc}    
\usepackage{hyperref}       
\usepackage{url}            
\usepackage{booktabs}       
\usepackage{amsfonts}       
\usepackage{nicefrac}       
\usepackage{microtype}      
\usepackage[pdftex]{graphicx}
\usepackage{amsmath,thmtools}
\usepackage{subfigure}
\usepackage{amsmath}
\usepackage{bm}
\usepackage{multirow}
\usepackage{float}
\usepackage[misc]{ifsym} 
\usepackage{color}
\usepackage[numbers]{natbib}
\usepackage{inputenc}
\usepackage{cleveref}
\usepackage{lineno}
\definecolor{grey}{rgb}{0.75, 0.75, 0.75}

\newcommand{\re}[1]{\textcolor{black}{#1}}

\title{EEGdenoiseNet: A benchmark dataset for \re{end-to-end} deep learning solutions of EEG denoising}
\author{Haoming Zhang$^{1,\dagger}$, Mingqi Zhao$^{1,2,\dagger}$, Chen Wei$^{1}$, \\Dante Mantini$^{2,3}$, Zherui Li$^{1}$, Quanying Liu$^{1,\ast}$}

\begin{document}
\maketitle
\vspace{-0.7cm}
\begin{centering}
$^1$ Shenzhen Key Laboratory of Smart Healthcare Engineering, Department of Biomedical Engineering, Southern University of Science and Technology, Shenzhen 518055, P.R. China\\
$^2$ Movement Control and Neuroplasticity Research Group, KU Leuven, Leuven 3001, Belgium \\
$^3$ Brain Imaging and Neural Dynamics Research Group, IRCCS San Camillo Hospital, Venice 30126, Italy \\
\vspace{0.5cm}
$^\dagger$  Equal contribution\\
$^\ast$ Corresponding author \texttt{liuqy@sustech.edu.cn} to Q.L.\\
\vspace{0.5cm}

\end{centering}
%
\section*{Abstract}
Deep learning networks are increasingly attracting attention in various fields, including electroencephalography (EEG) signal processing. These models provided comparable performance with that of traditional techniques. At present, however, lacks of well-structured and standardized datasets with specific benchmark limit the development of deep learning solutions for EEG denoising. Here, we present EEGdenoiseNet, a benchmark EEG dataset that is suited for training and testing deep learning-based denoising models, as well as for performance comparisons across models. EEGdenoiseNet contains 4514 clean EEG segments, 3400 ocular artifact segments and 5598 muscular artifact segments, allowing users to synthesize \re{contaminated} EEG segments with the ground-truth clean EEG. We used EEGdenoiseNet to evaluate denoising performance of four classical networks (a fully-connected network, a simple and a complex convolution network, and a recurrent neural network). Our analysis suggested that deep learning methods have great potential for EEG denoising even under high noise contamination. Through EEGdenoiseNet, we hope to accelerate the development of the emerging field of deep learning-based EEG denoising. \re{The dataset and code are available at \url{https://github.com/ncclabsustech/EEGdenoiseNet}}.\\ 

Keywords: Deep learning, Neural network, EEG dataset, Benchmark dataset, EEG artifact removal, EEG denoising

\section{Introduction}

Electroencephalography (EEG) solutions permit \re{recording of} changes in electrical potential on the scalp, which are generated by neurons in the gray matter. EEG is one of the most important direct and noninvasive approaches for studying brain activity under task and resting conditions. It has been widely used in \re{psychology, neurology and psychiatry} research, as well as for brain-computer interface \cite{cai2020feature-level,oberman2005eeg,wolpaw1991an,herrmann2005human,liu2017detecting,zhao2019hand}.\\

EEG signals contain not only brain activity, but also \re{a variety of} noise and artifacts, including ocular~\cite{croft2000removal}, myogenic artifacts~\cite{muthukumaraswamy2013high-frequency, piontonachini2019iclabel:}, and, in rare cases, cardiac artifacts. Therefore, a \re{basic} step in using EEG data to study neural activity is denoising or artifact attenuation~\cite{cohen2014analyzing}. Ocular and myogenic artifacts contaminate EEG signals in different ways. The former is often visible as relatively large pulses in the frontal region~\cite{chaumon2015a}, \re{while} the latter frequently appears in the  temporal and occipital regions, and has a wide frequency spectrum~\cite{piontonachini2019iclabel:, mcmenamin2010validation}.\\

Various traditional denoising techniques have been developed to remove artifacts from EEG data\re{, such as regression-based methods, adaptive filter-based methods and blind source separation (BSS)-based methods}. Among them, the regression-based method first obtains the noise signal through the noise template, and then subtracts the estimated noise signal from the EEG data to eliminate the artifacts \cite{mcmenamin2010validation, gratton1983new, croft2000eog, mcmenamin2009validation}. 
On the contrary, methods based on adaptive filters rely on dynamically estimating filter coefficients based on the input EEG signal itself, thereby filtering out noise~\cite{he2004removal, he2007removal}. 
BSS-based methods decompose the EEG signal into multiple components \cite{wold1987principal, hyvarinen2000independent, chen2019removal}, assign them to neural and artifactual sources, and reconstruct a clean signal by recombining the neural components~\cite{piontonachini2019iclabel:, mcmenamin2010validation, jung2000removal}. However, BSS-based methods can only be used when a large number of electrodes are available\re{, which are not suitable for single-channel denoising}.\\

Deep learning (DL) have been increasingly attracting attention in the past few years~\cite{krizhevsky2012imagenet, he2016deep, mikolov2013efficient, vaswani2017attention}. Due to the increase in  computing resources, the boosting data size, and the availability of new network architectures and learning algorithms, the performance of DL neural networks has made great breakthroughs, and deep learning has been successfully applied to solve various technical problems, such as image processing~\cite{krizhevsky2012imagenet, he2016deep, simonyan2014very, szegedy2016rethinking} and natural language processing \cite{mikolov2013efficient, vaswani2017attention, sutskever2014sequence}. 
DL methods have \re{begun to be introduced into the field of EEG signal analysis}~\cite{Roy_2019}, such as EEG-based classification \cite{2015Parallel, Yousef2017A, 2017Single}, EEG reconstruction \cite{2020Reconstruction, 2018Deep} and EEG signal generation \cite{2018generation, 2020Generative}. Recently, deep learning has also been applied to EEG denoising, providing performance comparable to the traditional denoising method~\cite{yang2018automatic, sun2020novel, Hanrahan2019NoiseRI,zhang2020novel}. \\

Deep neural networks can learn the hidden state of neural oscillations in EEG, thereby eliminating fluctuations that are not from real neural activity but from biological artifacts. The performance of deep neural networks fundamentally depends on the size of the training and \re{test} datasets; or in other words, it requires big data \cite{deng2009imagenet, defferrard2016fma, panchenko2017building}. A big dataset with the gold standard clean EEG is essential for evaluating newly developed supervised deep learning models. Some EEG datasets have been collected while participants are at rest \cite{trujillo2017effect,torkamani2020prediction}, during cognitive tasks \cite{koelstra2011deap,van2019building,korczowski2019brain}, or motor-related tasks \cite{luciw2014multi,HBM:HBM23730,schalk2004bci2000,kaya2018large}. \re{However, none of them are specifically developed for training end-to-end deep learning models for EEG artifact removal. To the best of our knowledge, there is no open EEG dataset suitable for training deep learning models for EEG denoising.} The lack of ground-truth clean EEG data and benchmarks have largely limited the development of DL methods for EEG denoising.\\

In this study, we present a publicly available structured dataset, named EEGdenoiseNet, which is particularly suitable for deep network-based EEG artifacts attenuation (\textbf{Sec} 2). Specifically, the dataset contains 4514 clean EEG segments as ground truth, and 3400 pure EOG segments and 5598 pure EMG segments as ocular artifacts and myogenic artifacts respectively. In addition, we also implement four deep neural networks as benchmarks (\textbf{Sec} 3), including a fully-connected neural network (FCNN), a simple convolution neural network (CNN), a complex CNN, and a recurrent neural network (RNN). \re{We train the deep learning models in a supervised end-to-end fashion, and the denosing performance are presented as benchmarks (\textbf{Sec} 4)}.\\



\section{EEGdenoiseNet Dataset}
\subsection{Data acquisition and preprocessing}
Our main goal is to \re{construct a dataset suitable for EEG denoising research based on deep learning networks}. In this regard, we downloaded EEG, EOG and EMG data from several publicly available data repositories which were published in previous studies (see Table \ref{table: data summary}) \cite{0EEG, 2016Assessing, 2007EMG, 2006Seperability, Schlogl2007A, 2007Spatial, article, inproceedings}. These studies have been ethically approved by their respective local ethical committees, and followed the Helsinki Declaration of 1975, revised in 2000. \\

\re{To generate clean EEG, pure EOG and pure EMG, we firstly preprocessed the data. Then segmented them into 2-second segments. Afterwards, we re-scaleed the segments to the same variance. Finally, each segment was visually checked by an expert to ensure they are clean and usable. We set the length of segments to 2 seconds according to the previous knowledge of EEG signals. On the one hand, a 2s segment is long enough to recover the temporal and spectral characteristics of EEG, as well as EOG and EMG. On the other hand, it is difficult to obtain artifact-free EEG segments longer than 2s due to the random eye blinks or movements. The segments in each category have been uploaded to a publicly available repository (\url{https://github.com/ncclabsustech/EEGdenoiseNet}).} \\

\begin{figure}[H]
\centerline{\includegraphics[width=\columnwidth]{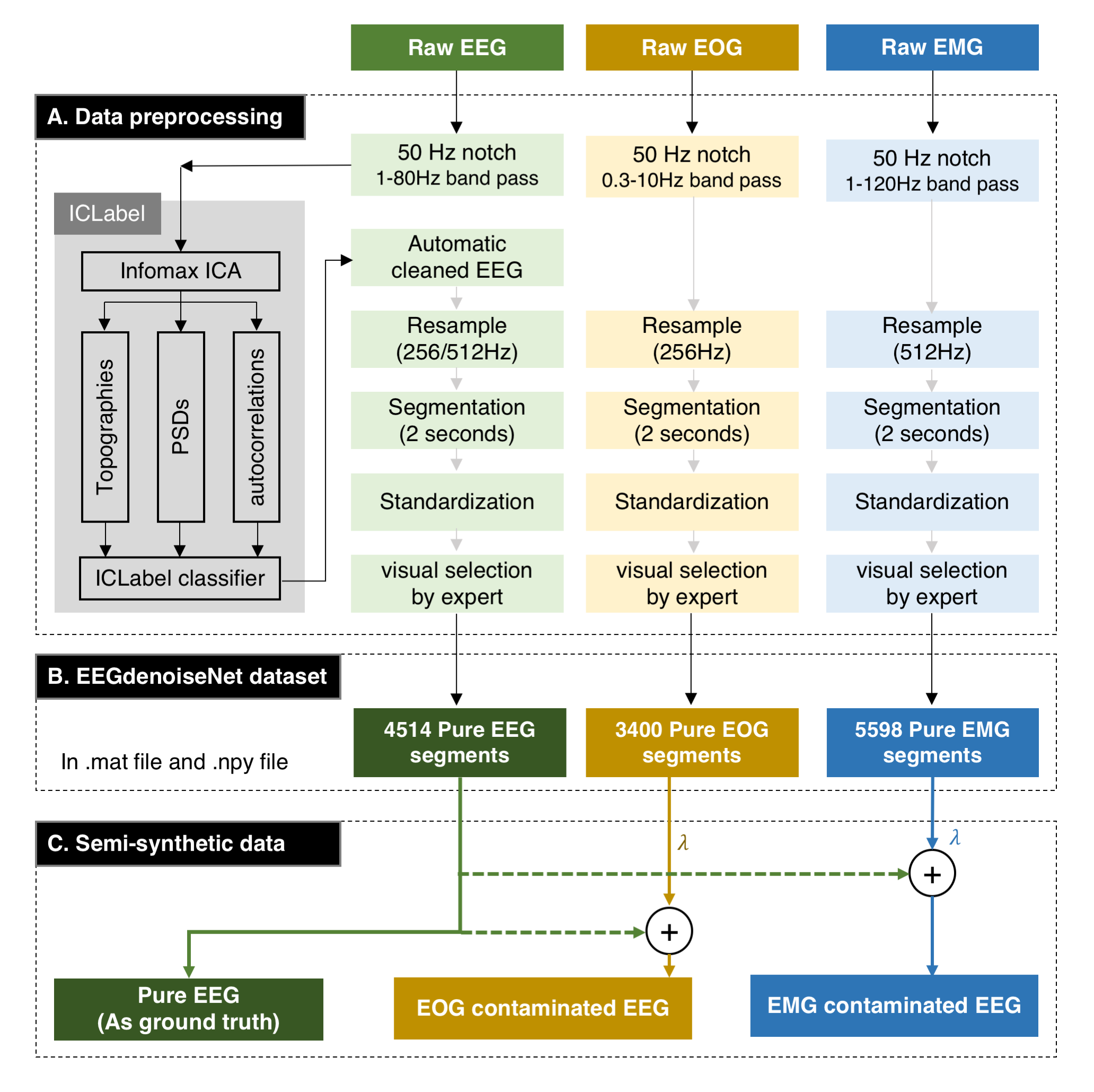}}
\caption{The pipeline for obtaining clean EEG, EOG and \re{EMG}. \re{(A) To obtain the clean EEG, pure EOG and pure EMG segments, we firstly preprocess the raw data. The data preprocessing include filtering, ICA-based artifacts removal, resampling, standarization, and visual checked by an expert. (B) 4514 pure EEG segments, 3400 pure EOG segments and 5598 pure EMG segments are obtained. EEGdenoiseNet dataset include two data formats: .mat files and .npy files. (C) The semi-synthetic data is generated by mixing a pure EEG segment and an EOG/EMG segment.}}
\label{fig:signal process}
\end{figure}

Specifically, for the EEG segments (Figure \ref{fig:signal process}$a$)\cite{0EEG}, \re{the dataset included 52 participants who performed both real and imaginary left and right hand movement task, with 64 channel EEG recorded simultaneously at 512 Hz sampling freqeuncy. For both real and imagined movement task, a participant repeated 2 second baseline and 3 seconds movement with 4.1 to 4.8 second random interval for 20 minutes.} The data was band-pass filtered between 1 to 80 Hz, notched at powerline frequency, and then re-sampled to 256 Hz. To obtain the clean EEG as ground truth, the 64-channel EEG signals were processed by ICLabel, a toolbox to remove EEG artifacts with independent component composition (ICA) \cite{piontonachini2019iclabel:}. Then the pure EEG signals were segmented into \re{one-dimensional} segments of 2 seconds. \re{It is worth noting that, in order to ensure the universality of this data set, we did not construct clean EEG signals with a specific number of channels due to the diversity of EEG caps, but constructed a dataset with single-channel EEG signal.}\\

For the ocular artifact segments (Figure \ref{fig:signal process}$b$), multiple open-access EEG datasets with additional EOG channels are used \cite{2016Assessing, 2007EMG, 2006Seperability, Schlogl2007A, 2007Spatial, article}. The horizontal and vertical \re{raw} electroculagraphy (EOG) signals of the datasets are band-pass filtered between 0.3 and 10 Hz, and then re-sampled to 256 Hz. Finally, the EOG signals are segmented into \re{one-dimensional} segments of 2 seconds.\\

For the myogenic artifact segments (Figure \ref{fig:signal process}$c$), a facial electromyography (EMG) dataset is used \cite{inproceedings}. We choose facial EMG because they are the main sources of myogenic artifacts. The \re{raw} EMG signal is band-pass filtered between 1 to 120 Hz and notched at the powerline frequency, and then resampled to 512 Hz. \re{We resample the EMG to 512 Hz instead of 256 Hz, because the EMG signal is concentrated in the high frequency range, so a higher sampling rate is required (according to the Nyquist sampling theorem)}. In the end, we extract \re{one-dimensional} 2-second EMG segments.\\

\begin{figure}[H]
\centerline{\includegraphics[width=\columnwidth]{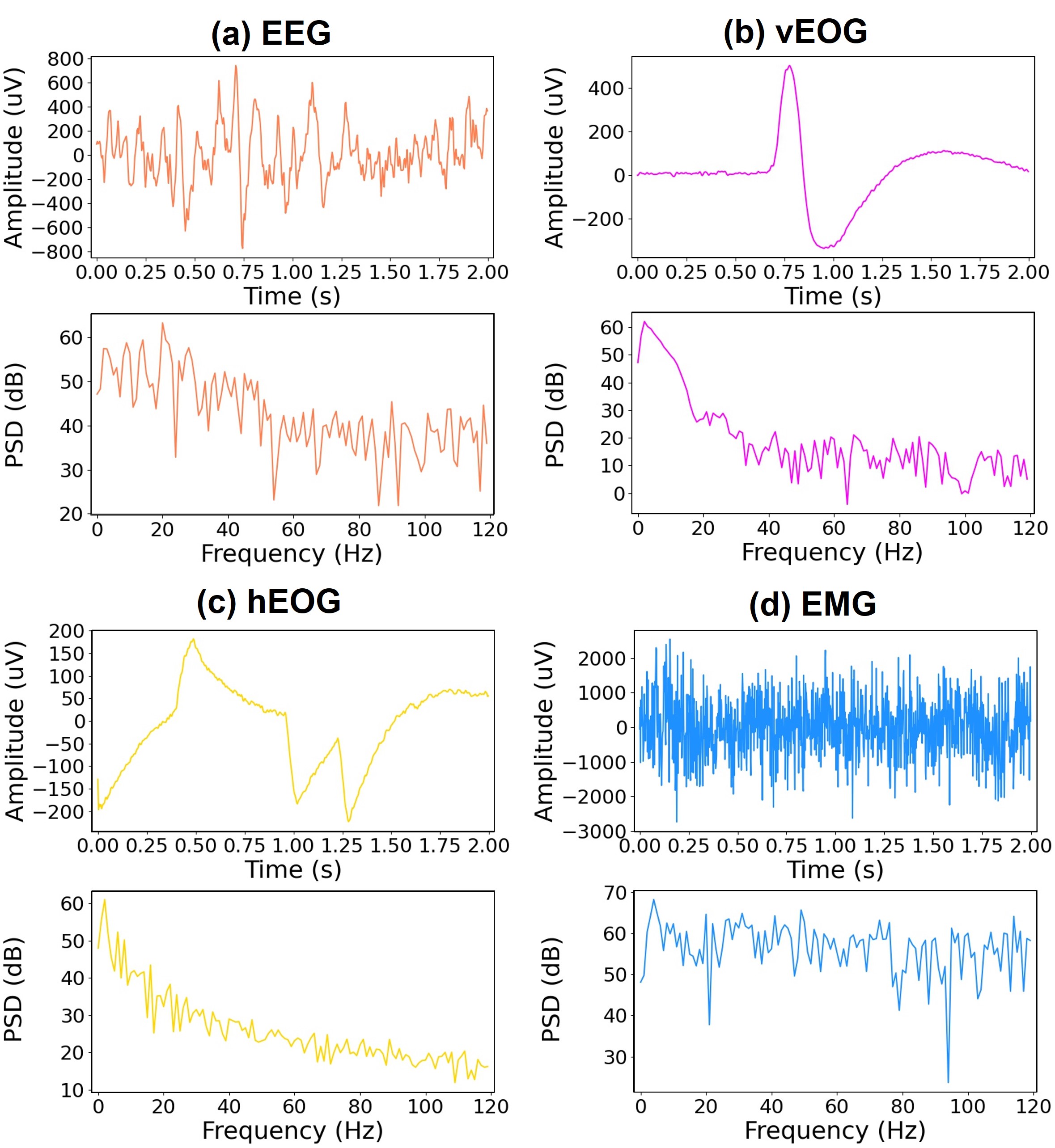}}
\caption{Examples of segments in EEGdenoiseNet dataset. (a) An EEG segment. (b) A vertical EOG (vEOG) segment. (c) A horizontal EOG (hEOG) segment. (d) An EMG segment. \re{(upper) The time course. (bottom) The PSD.} }
\label{fig:signal example}
\end{figure}

For all three categories, the segments are standardized by subtracting their mean and dividing by their standard deviation, and then visually inspected by an expert. We obtain a total of 4514 EEG segments, 3400 ocular artifact segments, and 5598 myogenic artifact segments. The segments of each category are respectively saved as Matlab matrix files and Python numpy matrix files in the public data repository. Figure \ref{fig:signal example} shows an example of the clean EEG, horizontal EOG, vertical EOG and EMG.

\subsection{Data Usage}
The \re{contaminated} signals can be generated by linearly mixing the pure EEG segments with EOG or EMG artifact segments, according to Eq. (\ref{Eq: EEG mixing}) (see Figure \ref{fig:signal process}$c$):

\begin{equation}
y = x +\lambda \cdot n\label{Eq: EEG mixing}
\end{equation}
where $y$ denotes the mixed one-dimensional signal of EEG and artifacts; $x$ denotes the clean EEG signal as the ground truth; $n$ denotes (ocular or myogenic) artifacts; \re{$\lambda$ is a hyperparameter to control the signal-to-noise ratio (SNR) in the contaminated EEG signal $y$.} Specifically, the SNR of the \re{contaminated} segment can be adjusted by changing the parameter $\lambda$ according to Eq. (\ref{Eq: SNR}):

\begin{equation}
SNR =10\log {\frac{RMS(x)}{RMS(\lambda \cdot n)}}\label{Eq: SNR}
\end{equation}
in which the Root Mean Squared (RMS) value is defined as Eq. {\ref{Eq:RMS}}:
\begin{equation}
RMS(g) = \sqrt{\frac1{N}\sum_{i=1}^Ng_i^2}\label{Eq:RMS}
\end{equation}
where $N$ denotes the number of temporal samples in the \re{segment} $g$, and $g_i$ denotes the $i^{th}$ sample of a segment $g$. \re{Notably, lower $\lambda$ represents higher SNR, as less EOG or EMG artifacts are added in the EEG signal. In return, lower SNR means higher noise level. According to previous studies, the SNR of EEG contaminated by ocular artifacts is usually ranging from -7dB to 2dB~\cite{2016The}, while the SNR of EEG contaminated by myogenic artifacts are between -7dB and 4dB~\cite{2017Independent, 2006Canonical}.} \\

\re{In this way, we obtain a pair of EEG data ($x,y$). To train the end-to-end deep learning methods for EEG denoising, the clean EEG $x$ can be regarded as the ground truth, and the contaminated EEG $y$ can be used as the inputs.}

\section{Benchmarking deep learning algorithms}
The second goal of this study is to provide a set of benchmark algorithms. We train four standard deep-learning neural networks, then validate \re{the networks}. The evaluation metrics can be used as benchmarks for the EEG denoising algorithms.

\subsection{Generating semi-synthetic data}
\re{The semi-synthetic ocular artifact contaminated signals are from 3400 EEG segments and 3400 ocular artifact segments, with 80\% for generating the training set, 10\% for generating the validation set, and 10\% for generating the test set \cite{1995A}. Each set were generated by randomly linearly mixing EEG segments and ocular artifact segments according to section 2.2, with SNR raging from ten different SNR levels (-7dB, -6dB, -5dB, -4dB, -3dB, -2dB, -1dB, 0dB, 1dB, 2dB). This procedure expanded the data size of each set to ten times. The EEG segments are treated as ground truth, and the corresponding mixed segments are treated as contaminated EEG.}\\

\re{The myogenic artifacts contaminated signals come from 4514 EEG segments and 5598 myogenic artifact segments. To match the sampling frequency of EEG segments with myogenic artifact segments, we upsample the EEG segments to 512Hz. To match the number of EEG segments with myogenic artifact segments, we randomly reuse some EEG segments. We mix the EEG segments and the myogenic artifact segments as Eq.(\ref{Eq: EEG mixing}) to generate the training data, test data, and validation 
data. Likewise, the EEG segments are treated as ground truth, and the corresponding mixed segments are treated as contaminated EEG}.

\subsection{Network architectures}
\subsubsection{Fully-connected Neural Network}
\begin{figure}[H]
\centerline{\includegraphics[width=\textwidth]{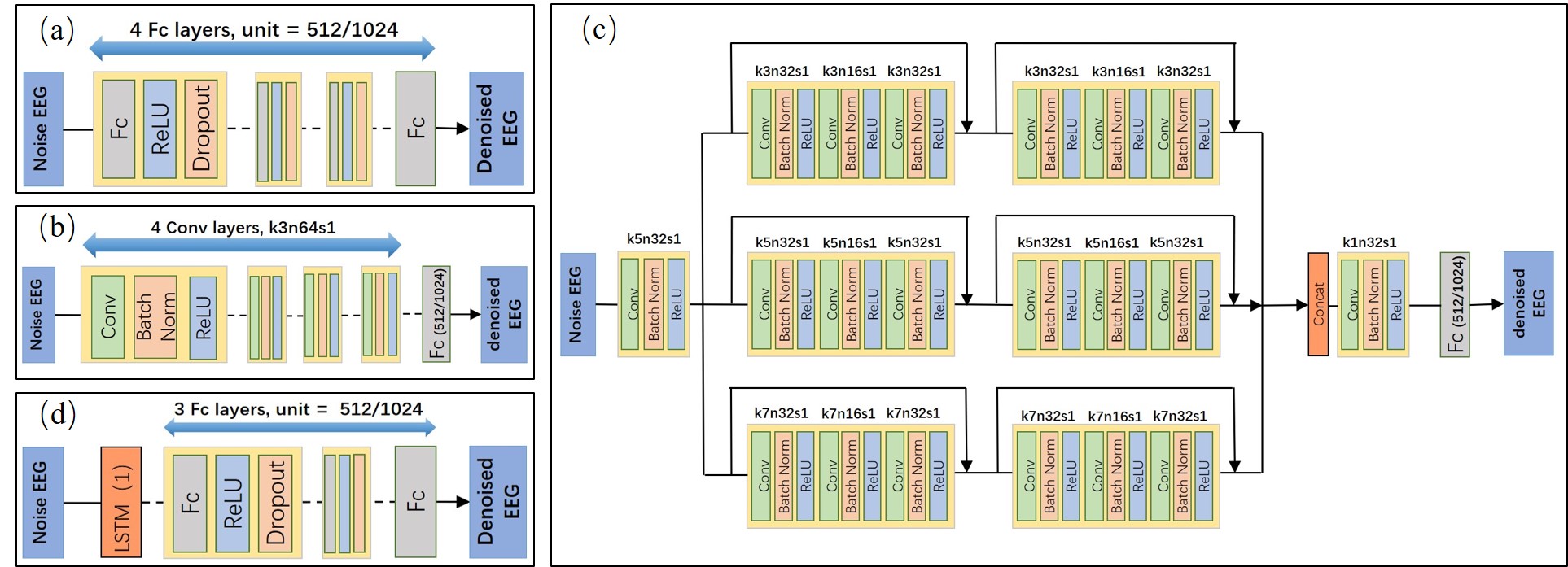}}
\caption{The structures of the four DL-based methods for benchmarking. (a)FCNN ; (b) Simple CNN; (c) Complex CNN; (d) RNN. \re{For each of the networks, the input is the contaminated segments ($1\times512$ for ocular artifact removal, and $1\times1024$ for myogenic artifact removal), and the output is the cleaned segments ($1\times512$ for ocular artifact removal, and $1\times1024$ for myogenic artifact removal). The networks are trained to learn the genuine neural activities from contaminated signal segments according to contaminated segments and ground truth segments.}}
\label{fig:network-structure}
\end{figure}

A fully-connected network with four hidden layers using ReLu as activation function is provided as a benchmarking algorithm (Figure \ref{fig:network-structure}a). The number of neurons in each layer is equal to the number of temporal samples of the input signal (\textit{i.e.}, 512 for ocular artifact reduction, and 1024 for myogenic artifact reduction). Dropout regularization \cite{hinton2012improving} is used to reduce overfitting. \re{The contaminated EEG is fed in from the first layer of the neural network, and then the denoised EEG is output from the last layer.}

\subsubsection{Simple Convolution Neural Network}
A simple convolution network is implemented (Figure \ref{fig:network-structure}b). The simple CNN consists of four 1D-convolution layers with small 1$\times$3 kernels, 1 stride, and 64 feature maps (k3n64s1). Each 1D-convolution layer is followed by a batch-normalization (BN) layer~\cite{ioffe2015batch} and a ReLu activation function. To reconstruct the signal, the last convolutional layer is followed by a flatten layer and a dense layer with 512 or 1024 neurons as outputs (the same as the input).

\subsubsection{Complex Convolution Neural Network}
An one-dimensional residual convolutional neural network (1D-ResCNN), adapted from \cite{sun2020novel}, is implemented (Figure \ref{fig:network-structure}c). \re{Compared with simple CNN, the 1D-ResCNN has a more complex structure, so it is called complex CNN}. The main difference between them is that a ResNet with skip-layer connections is added into the complex CNN to avoid gradient explosion so that we can train a deeper network to obtain better feature extraction capabilities \cite{he2016deep}. To extract multi-scale features, we repeatedly stack residual blocks, using 1$\times$3, 1$\times$5, 1$\times$7 multi-scale convolutional kernels twice and arranging three sets of residual blocks branches in parallel \cite{szegedy2016rethinking, zagoruyko2016wide}.

\subsubsection{Recurrent Neural network}
A Long Short-Term Memory (LSTM) network (Figure \ref{fig:network-structure}d), adapted from
\cite{hochreiter1997long}, is regard as the benchmark of recurrent neural networks (RNNs). LSTM can learn long-term dependencies, which may help distinguish the long-term features in noise and EEG signals. Each EEG sample is sequentially input to LSTM cells, and the output is obtained from the state of each cell through a fully-connected network. This RNN model is initialised to have 1 hidden state, and the output network is a three-layer fully-connected network with ReLu activation function, dropout regularization, and 512 or 1024 neurons per layer.

\subsection{Learning process}
In order to facilitate the learning procedure, we normalized the input \re{contaminated} EEG \re{segment} and the ground-truth EEG  \re{segment} by dividing the standard deviation of \re{contaminated} EEG \re{segment} according to Eq. (\ref{Eq:normalization}):

\begin{align}
\label{Eq:normalization}
\hat{x} = \frac{x}{\sigma_{y}},  &  &  \hat{y} = \frac{y}{\sigma_{y}}
\end{align}
where $\sigma_{y}$ is the standard deviation of the \re{contaminated} EEG signal \re{segment} $y$. The standard deviation of each noise \re{segment} is saved, so that the magnitude of the denoised EEG \re{segment} can be restored by multiplying the network output by the corresponding standard deviation value. \\

\re{The networks are trained in an end-to-end manner, which means that we input the normalized contaminated EEG segment into the neural networks and then directly output the denoised EEG segment.}  To this end, the goal of a denoising network is to learn a nonlinear function $f$ that maps the \re{contaminated} EEG $\hat{y}$ to the \re{denoised} EEG $\tilde{x}$:
\begin{equation}
\tilde{x} = f(\hat{y}, \boldsymbol{\theta} )
\label{Eq: map function}
\end{equation}
where $\hat{y}\in \mathbb{R}^{1\times T}$ denotes the \re{contaminated} EEG \re{segment}, $\tilde{x}\in \mathbb{R}^{1\times T}$ as the output of neural network (the \re{denoised} EEG \re{segment}), and the vector $\boldsymbol{\theta}$ contains all parameters to be learned. \\ 

We use the mean squared error (MSE) as loss function $L_{MSE}(f)$ (see Eq. (\ref{Eq: loss})). The learning process is implemented with gradient descent to minimize the error between the denoised \re{segment} and the ground-truth clean \re{segment}.

\begin{equation}
L_{MSE}=\frac1{N}\sum_{i=1}^N\Big|\Big|\tilde{x}_i-\hat{x}_i\Big|\Big|^2_2\label{Eq: loss}
\end{equation}
where $N$ denotes the number of temporal samples of \re{segment}; $\tilde{x}_i$ denotes $i^{th}$ sample of the output of the neural network; $\hat{x}_i$ denotes the $i^{th}$ sample of the ground truth $x$.\\

For ocular artifact removal, we train the FCNN with 60 epochs, RNN with 100 epochs, while the simple CNN and complex CNN models are trained over 40 epochs. For myogenic artifact removal, we train the FCNN and RNN with 60 epochs, while the simple CNN and complex CNN models are trained over 10 epochs. The Adam algorithm is applied to optimize the loss function, and the parameter were set to $\alpha = 5e^{-5}$, $\beta_1=0.5$, $\beta_2=0.9$. To increase the statistical power, the four networks are trained, validated and tested independently for 10 times with randomly generated datasets via EEGdenoiseNET.\\

All the four networks are implemented in Python 3.7 with Tensorflow 2.2 library, running on a computer with two NVIDIA Tesla V100 GPUs. The codes for the benchmarking algorithms are publicly available online at Github~\cite{EEGdenoiseNET}.

\subsection{Performance Evaluation as Benchmarks}
\re{There are several metrics are used to qualitatively evaluate the performance of networks, including the network convergence, the relative root mean squared error, and the correlation coefficient. } \\

The network convergence is the first index to evaluate the performance of networks, which can provide rich information about the learning procedure and generalization ability. The convergence curve of both training and validating processes are obtained by calculating the averaged loss (in Eq. (\ref{Eq: loss})) with respect to the number of epochs.\\

We then quantitatively examine the performance of the networks by applying three objective measures \re{to} the denoised data \cite{2017Independent}, including Relative Root Mean Squared Error (RRMSE) in the temporal domain ($RRMSE_{temporal}$, see Eq. (\ref{Eq: RRMSEt}) ), RRMSE in the spectral domain ($RRMSE_{spectral}$, see Eq. (\ref{Eq: RRMSEf}) ) and the correlation coefficient ($CC$ see Eq. (\ref{Eq: CC})).
\begin{equation}
RRMSE_{temporal} =\frac{RMS(f( y ) -x)}{RMS(x)}\label{Eq: RRMSEt}
\end{equation}

\begin{equation}
RRMSE_{spectral} =\frac{RMS(PSD(f(y))-PSD(x))}{RMS(PSD(x))}\label{Eq: RRMSEf}
\end{equation}
where the function $PSD( )$ denotes to the power spectral density of an input \re{segment}. \re{The frequency range of $PSD$ is 0-120Hz. The fft-length equal to the length of the input segment.}

\begin{equation}
CC =\frac{Cov(f(y),x)}{\sqrt{Var(f(y))Var(x)}}\label{Eq: CC}
\end{equation}


\re{To compare the deep learning mehtods with the traditional methods,} we implement two traditional EEG denoising methods: i) empirical mode decomposition (EMD) and ii) filtering. In the EMD method, the artifactual intrinsic mode functions (IMFs) are defined by the distance metric used in clustering \cite{2014Hilbert}. In the filtering method, the ocular and myogenic artifacts are removed using a high-pass filter (12 Hz) and a band-pass filter (12-40 Hz) , respectively. \re{These two traditional methods are tested 10 times with randomly generated datasets.} The corresponding codes are available online at \url{https://github.com/ncclabsustech/Single-Channel-EEG-Denoise}.

\section{Results}
\begin{figure}[H]
\centerline{\includegraphics[width=0.85\columnwidth]{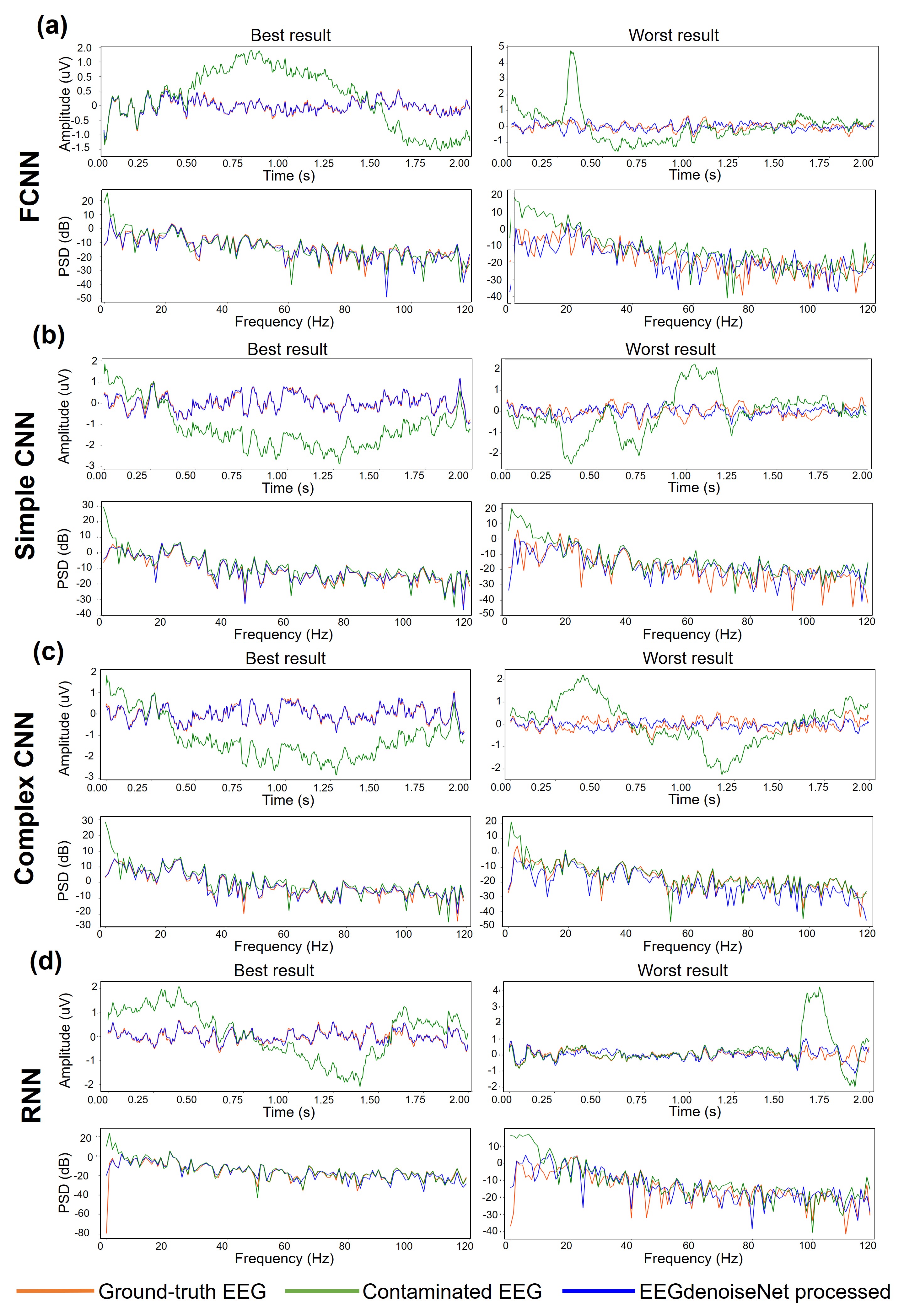}}
\caption{Some examplary segments of the performance in temporal domain (upper) and spectral domain (bottom) for ocular artifact removal. (a) FCNN. (b) Simple CNN. (c) Complex CNN. (d) RNN. (left) The examples with the best denoising performance; (right) the examples with the worst denosing performance. The orange, green and blue line are the ground-true EEG, the noisy EEG and the cleaned EEG by EEGdenoiseNet, respectively.}
\label{fig:EOG_results_examples}
\end{figure}

\begin{figure}[H]
\centerline{\includegraphics[width=0.85\columnwidth]{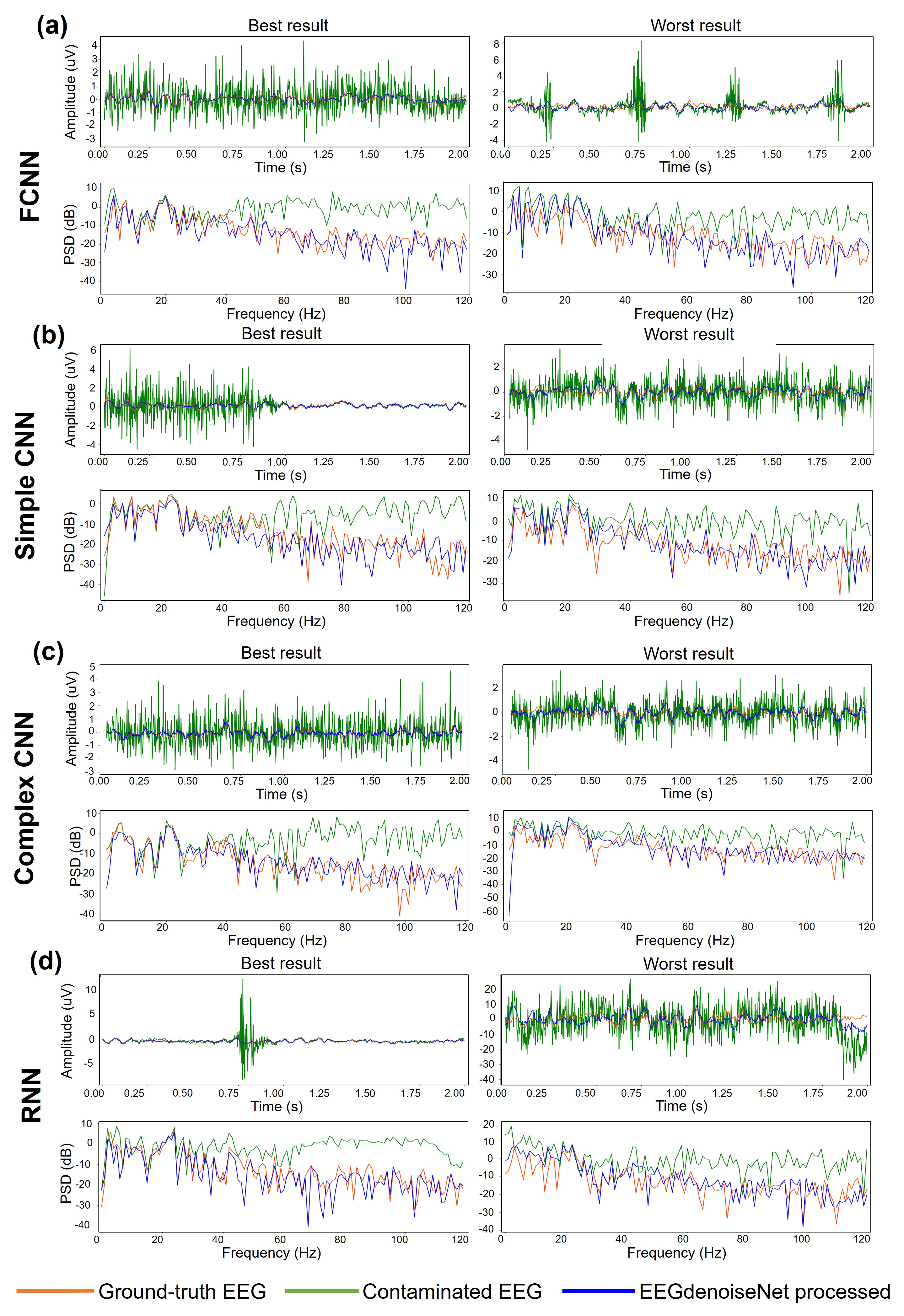}}
\caption{Some examplary segments showing the performance in temporal (\textit{upper}) and spectral (\textit{bottom}) domains for myogenic artifact removal. (a) FCNN. (b) Simple CNN. (c) Complex CNN. (d) RNN. (\textit{left}) The examples with the best denoising performance; (\textit{right}) the examples with the worst denoising performance. The orange, green and blue lines are the ground-true EEG, the noisy EEG and the cleaned EEG by EEGdenoiseNet, respectively.}
\label{fig:EMG_results_examples}
\end{figure}

To give a qualitative overview of the denoising results, we display some \re{sample fragments} in the test in the time domain and frequency domain for ocular artifact removal (see Figure \ref{fig:EOG_results_examples}) and for myogenic artifact removal (see Figure \ref{fig:EMG_results_examples}). For each network and artifact type, we show two examples: one of the best results (left column) and one of the worst result (right column). Generally, both in ocular and myogenic artifact removal, the artefacts are greatly attenuated, and the noise-free EEG samples are well-reconstructed. The frequency domain diagram shows that the artifacts in the low frequency bands are well detected and attenuated, but the high frequency is affected by residual noise.\\

\begin{figure}[H]
\centerline{\includegraphics[width=1\columnwidth]{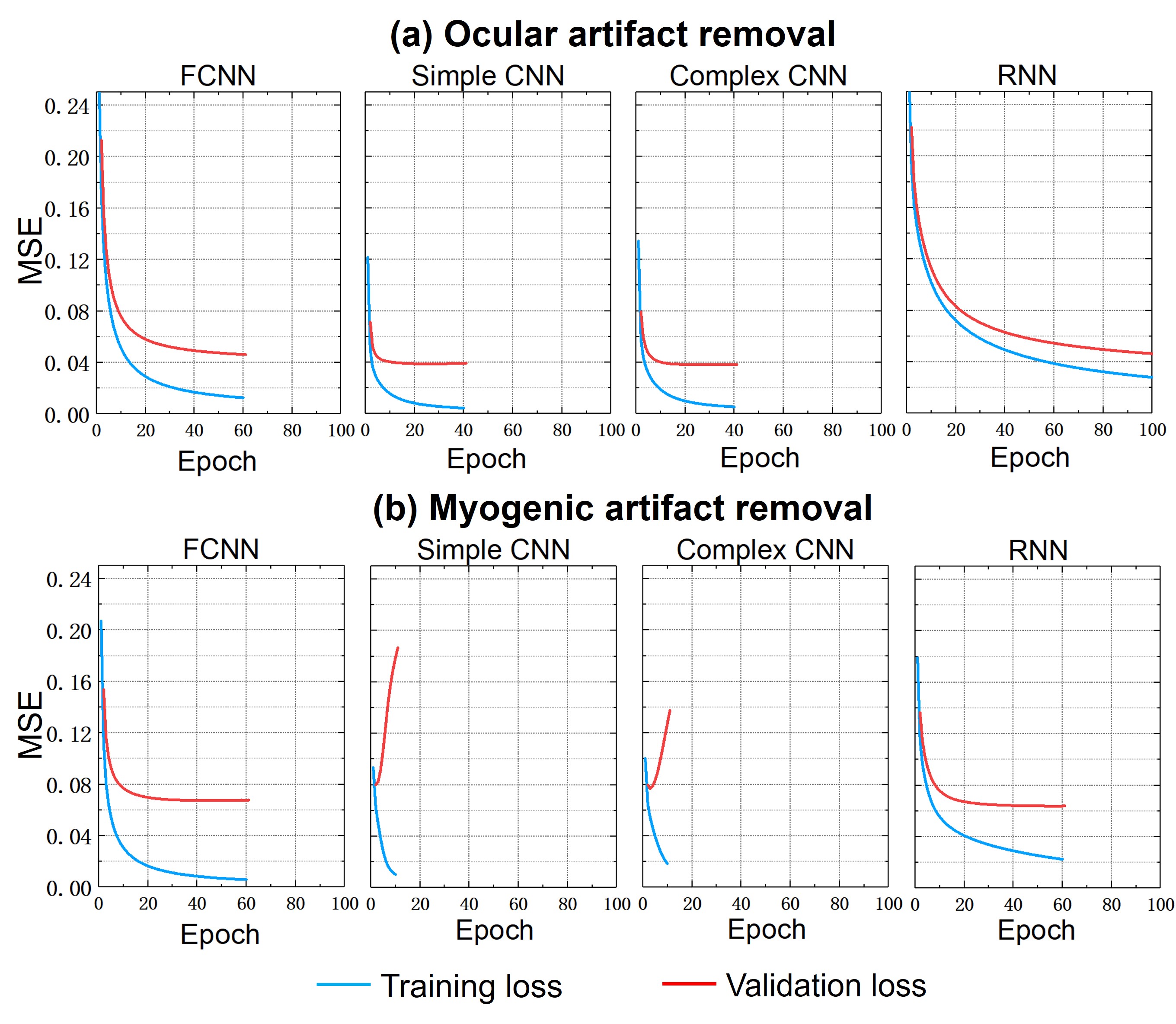}}
\caption{The MSE loss as a function of the number of epochs: (a) ocular artifact removal; (b) myogenic artifact removal. \re{The red line is the learning curve for the training set and the blue line for the validation set. } }
\label{fig:EOG_EMG_loss}
\end{figure}

The quantitatively results are examined. We first present the convergence of the four networks. The training and validation loss of the networks can show a quantitative overview of the training and validating process. For all the 4 networks and 2 artifact types, the training loss is consistently lower than the validation loss as expected. For the ocular artifact removal (see Figure \ref{fig:EOG_EMG_loss}a), the training and validation loss decrease with the increase of epochs. Specifically, the loss of simple CNN and complex CNN drop notably fast and eventually diminish after 20 epochs. The FCNN loss and the RNN loss, however, starting from a relatively high level, remain at a significant level after 20 epochs. For the myogenic artifact removal (see Figure \ref{fig:EOG_EMG_loss}b), the training loss of four networks decreased with respect to the number of epochs, similar to the ocular artifact removal. The loss of simple CNN and complex CNN decrease \re{faster} during training, but increased during validation. \re{This phenomenon indicates that the two convolutional networks seem not learn the true characteristics of the EMG signal, which means that CNNs suffer from an overfitting problem when removing myogenic artifacts.} \\

\begin{figure}[H]
\centerline{\includegraphics[width=1\columnwidth]{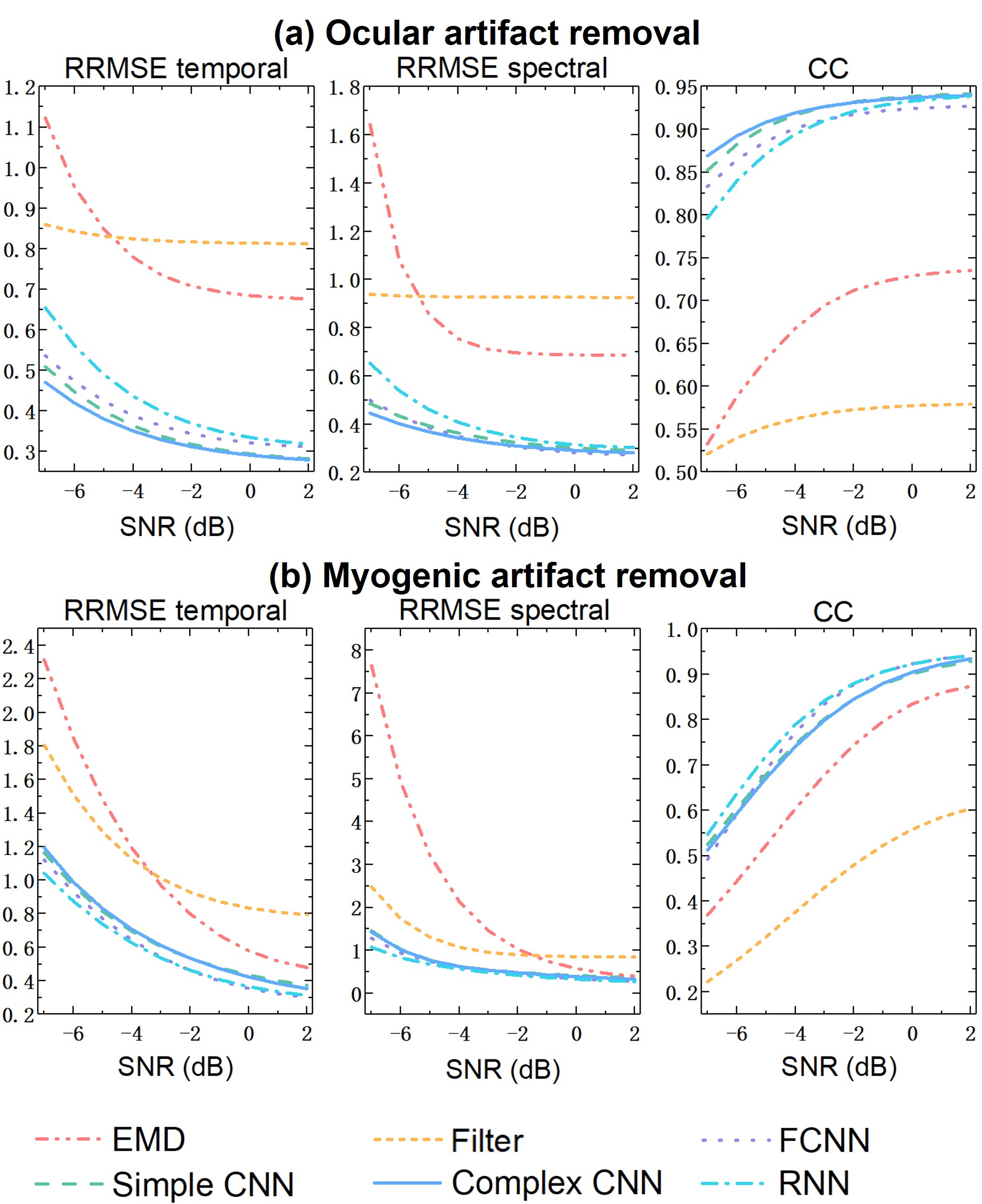}}
\caption{Performance of four deep-learning networks at different SNR levels: (a) ocular artifact removal; (b) myogenic artifact removal. \re{The cold colored lines represent deep learning methods, while the warm colored lines represent traditional methods. The denoising performance increases as the SNR increases.}}
\label{fig:evaluation}
\end{figure}

We present the quantitative benchmarks ($RRMSE_{temporal}$, $RRMSE_{spectral}$ and $CC$) from the four networks and the two traditional methods at multiple SNR levels (see Figure \ref{fig:evaluation}). Generally, for both ocular and myogenic artifact removal, the performance decreases with the decrease of SNR level. The traditional methods showed higher $RRMSE$ and lower $CC$ compared with the four deep learning networks. The difference of performance is larger at the large noise level (low SNR), while the difference reduces at low noise level (large SNR, eg. $SNR > 0$). Among the deep learning methods, RNN has the lowest $RRMSE$ and the highest $CC$ for ocular artifact removal (see Figure \ref{fig:evaluation}a), and the complex CNN has the lowest $RRMSE$ and the highest $CC$ in myogenic artifact removal.\\

\begin{figure}[H]
\centerline{\includegraphics[width=1\columnwidth]{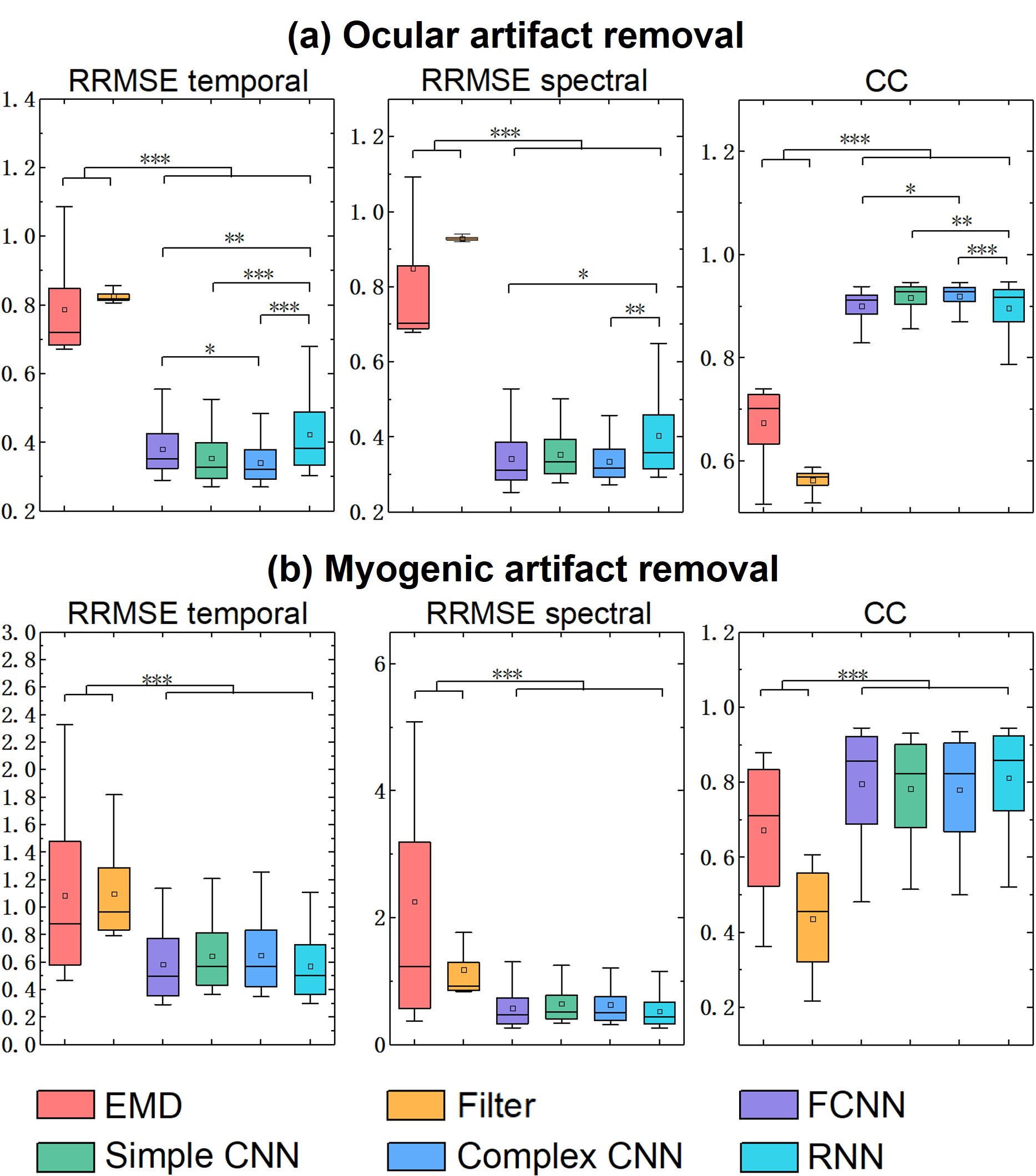}}
\caption{\re{Performance of four DL networks (FCNN, simple CNN, complex CNN, RNN) and two traditional methods (EMD and filter): (a) Ocular artifact removal, (b) Myogenic artifact removal. Deep learning models robustly outperform EMD and filtering for EEG denosing.}}
\label{fig:boxplot}
\end{figure}

To further comprehensively compare benchmarks, we separately plot the benchmarks at multiple SNR levels in boxplot (see Figure \ref{fig:boxplot}), and conduct ANOVA analyses. For the ocular artifact removal (see Figure \ref{fig:boxplot}a),  DL-based methods have significantly better denoising performance compared to two traditional methods, in terms of $RRMSE_{temporal}$, $RRMSE_{spectral}$ and $CC$ ($p<0.001$ for each of three metrics). Similarly, the DL-based methods outperform traditional methods for myogenic artifact removal ($p<0.001$ for each of three metrics) (see Figure \ref{fig:boxplot}b). In the time domain,  $RRMSE_{temporal}$ of RNN is significantly higher than FCNN, simple CNN, and complex CNN ($p=0.007$, $p<0.001$, and $p<0.001$, respectively); FCNN has significantly higher $RRMSE_{temporal}$ than the complex CNN ($p=0.020$). At the frequency domain,  RNN has significantly higher $RRMSE_{spectral}$ than the FCNN and the complex CNN ($p=0.020$ and $p=0.006$, respectively). $CC$ of the complex CNN is significantly higher than RNN and FCNN ($p<0.001$, and $p=0.011$, respectively). The same effect is shown on simple CNN and RNN ($p=0.004$).\\

We finally evaluate the performance of the different methods for different frequency bands, by calculating the average power ratio of each frequency band (delta [1-4 Hz], theta [4-8 Hz], alpha [8-13 Hz], beta [13-30 Hz], and gamma [30-80 Hz] bands) to whole band (1-80 Hz) for ocular artifact removal and myogenic artifact removal \re{(see Table \ref{table: power of pre/post_EEG} and \ref{table: power of pre/post_EMG})}. For the ocular artifact removal (see Table \ref{table: power of pre/post_EEG}), the mix of ocular artifacts increased the power ratio of delta and theta bands, while reduced the ratio of the other bands. The simple CNN showed the closest delta, theta and beta power ratio compared to those of the ground truth, the same effect observed on the complex CNN for theta band, on the EMD for alpha band, on the RNN for gamma band, and on the FCNN for theta and gamma band. In myogenic artifact removal (see Table \ref{table: power of pre/post_EMG}), the add of myogenic artifact increased gamma power ratio and decreased other power ratios. The FCNN showed the closest ratio in beta bands compared to those of the ground truth, the same effect on the EMD for alpha band, on the Complex CNN for theta and gamma bands, on the RNN for delta and alpha bands.

\section{Discussion}
In this study, we have provided an EEG benchmark dataset, EEGdenoiseNet, for training and testing end-to-end deep learning models. To obtain the ground-truth clean EEG data, the raw EEG data is denoised by ICLabel \cite{piontonachini2019iclabel:} and then manually inspected for a double check. \re{Although there are other publicly available EEG datasets, they are not specifically developed for EEG denoising. Instead, they are mainly focused on the resting state study \cite{trujillo2017effect,torkamani2020prediction}, psychological study \cite{koelstra2011deap,van2019building,korczowski2019brain}, or motor imaginary or motor tasks \cite{luciw2014multi,HBM:HBM23730,schalk2004bci2000,kaya2018large}.} A previous study has offered a semi-simulated dataset for EOG artifact removal, but EMG signals are not included \cite{klados2016semi}. Effective use of these datasets for DL-based denoising requires extensive EEG background knowledge, including properties of EEG and artifacts, data format conversion, and signal processing. In contrast, the segments in our dataset have been pre-processed, so users can immediately generate a large set of semi-synthetic noisy EEG segments with ground truth for their DL-based denoising models without being distracted by detailed electrophysiological knowledge. With this advantage, our well-structured dataset would greatly promote the development of DL-based EEG denoising.\\

Another major challenge to compare the performance of different denoising algorithms is the lack of specific benchmarks. The use of standard benchmarks greatly simplify the comparisons of performance across multiple DL models. To fill this gap, we provided a set of benchmark algorithms along with a standardized EEG dataset. We chose four well-known and relatively basic networks, i.e. a FCNN, a simple CNN,  and a RNN for benchmarking. Performance of these DL models in providing artifact-corrected EEG data has been measured using several standard metrics, such as RRMSE, PSD and CC. Furthermore, we define the network convergence, expressed by loss as a function of epoch number, as a qualitative part of the benchmarks. We expect our work to contribute to the DL-based EEG denoising field, in particular for we standardizes evaluation metrics of performance.\\

Our benchmarks of four deep learning networks and two traditional methods have demonstrated the feasibility of using DL-based methods to attenuate artifacts from EEG signals. Our comparisons of the four networks (\textit{i.e.}, FCNN, simple CNN, complex CNN, RNN) with two traditional methods (\textit{i.e.}, EMD and filter) suggest that DL-based methods outperform the traditional method, and the supervised end-to-end deep learning has great potential to remove artifacts in EEG signals. Specifically, for the ocular artifacts,  the range of $CC$ values in our four networks are at equivalent level of the $CC$ values reported in a previous study, which used a regression-based method and an offline ICA-based method \cite{guarnieri2018online}. A consistent result has been also reported by a DL-based ocular artifact removal study~\cite{yang2018automatic}. However, these studies have not offered benchmarks for comparing with other methods \cite{nguyen2012eog, woestenburg1983removal, elbert1985removal, gomez2006automatic}. For the myogenic artifacts, comparable $RRMSE_{temporal}$ values have been reported in previous literature, such as an ICA-based method \cite{2006Canonical} and a canonical correlation analysis-based method \cite{chen2017independent}. \\

\re{The performance of the neural networks depends on the data quality and frequency characteristics of artifacts.} The neural networks provide better results for high SNR signals than low SNR signals (Figure \ref{fig:evaluation}). Moreover, the high-frequency artifacts, such as EMG artifacts, are more difficult for neural networks to deal with (Figure \ref{fig:evaluation}-\ref{fig:boxplot}). This phenomenon may be explained by the \textit{F-Principle} of neural network \cite{xu2019frequency}. The F-Principle proves that deep learning networks often learn low-frequency information in the early stages of training, and then learn high-frequency information as training iterations increase.\\

One advantage of deep learning for EEG artifact removal is its flexibility and generalizability. \re{Although the} DL-based denoising methods require a large amount of ground-truth EEG data in the training stage, once the model is trained, it can be easily applied to new data, such as multi-channel EEG data or task-related EEG data, regardless of the corresponding reference channels for artifact removal. Another advantage lies \re{in the handling of} complex (\textit{e.g.}, nonlinear and non-stationary) artifact mixtures. Due to the hierarchical structure of deep neural networks, DL models can directly learn the true nature of neural activities from training data in the hidden space, and then generate the cleaned EEG data according to the new \re{contaminated} EEG input, whereas traditional methods usually linearly attenuate artifacts. Therefore, methods based on deep learning are expected to provide better performance than traditional methods in noise removal.\\

Several limitations should also be mentioned. \re{First of all,} an important potential problem is the size of the dataset and the type of data. Although we provided thousands of segments of EEG, ocular and myogenic artifacts in EEGdenoiseNet, it is possible that more complex neural networks might require larger amounts of data for training and testing. Another drawback is the diversity of the EEG type and artifact type. EEG data may be collected in resting state or in different task conditions; furthermore, artifacts in EEG recordings are not only limited to ocular and myogenic. For example, \re{removing motion artifacts is important for EEG mobile applications}. \re{Criteria} for reviewing and approving additional EEG data submissions to EEGdenoiseNet would be helpful. Such an evolving dataset will contribute to improve the generalibility of the DL-based EEG denoising networks to diverse brain states. Third, we only focused on the denosing of 2s-long EEG segments in this study. It is worth noting that some EEG tasks might be longer than 2 seconds, not to mention the case of resting EEG. In the future, EEGdenoiseNet needs to be extended to adapt to the denoising of continuous EEG. The continuous artifact removal problem can be solved by defining pseudo-segments in continuous EEG recordings, and extracting the hidden relationships between consecutive segments, such that the previous segment can be used in the training stage as input to constrain the denoising process of the current segment. \re{Forth, here we only focus on single-channel EEG denoising, and the deep learning model learns the temporal information of EEG signals and EOG/EMG artifacts. To use supervised models to learn spatial features, a benchmark data set with multi-channel EEG data must be provided in future.} \re{Finally, we did not explore unsupervised deep learning models in this study. When there is no gold standard for clean EEG signals and artifacts, unsupervised deep learning may be of great importance.}

\section{Conclusion}
In this study, we provided a dataset containing thousands of clean EEG, ocular artifact\re{s} and muscular artifact segments, which is suited for benchmarking DL-based EEG denoising methods. The dataset is well-structured and publicly available online in different formats. In addition, we included a set of benchmark tools to facilitate the evaluation of newly developed DL-based EEG denoising models. Our benchmarking results suggested that DL methods have great potential to remove both ocular and myogenic artifacts from EEG data, even at high noise levels. Our study may accelerate the development of DL-based EEG denoising field.

\section*{Acknowledgement}
This work was funded in part by the National Natural Science Foundation of China (62001205), Guangdong Natural Science Foundation Joint Fund (2019A1515111038), Shenzhen Science and Technology Innovation Committee (20200925155957004, KCXFZ2020122117340001, \\ SGDX2020110309280100), Shenzhen Key Laboratory of Smart Healthcare Engineering \\ (ZDSYS20200811144003009).

\section*{Competing intersts}
The authors declare no competing interests.

\bibliographystyle{unsrt}

\newpage
\section*{Figures and Tables}

\newpage

\newpage
\begin{figure}[H]
\centerline{\includegraphics[width=\columnwidth]{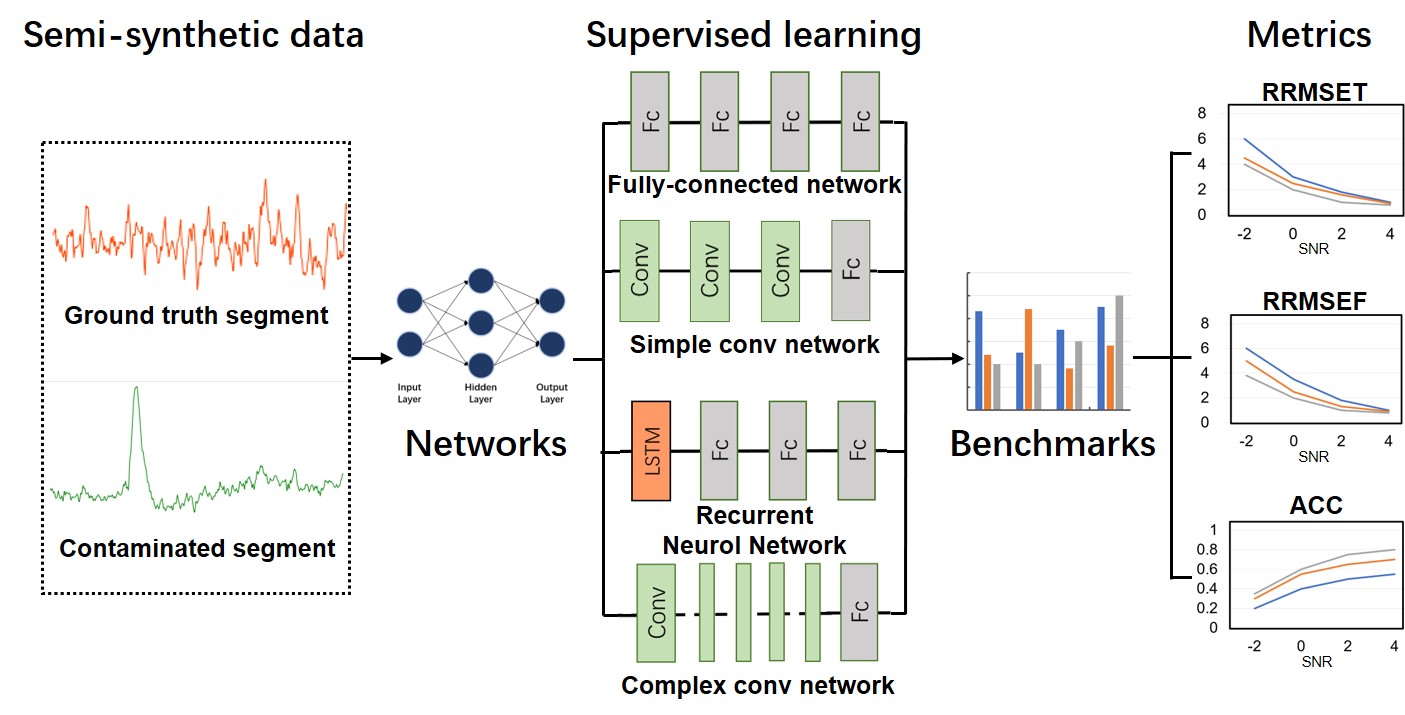}}
\caption{The framework of EEGdenoiseNet. \re{The semi-synthetic data includes ground truth EEG segment and contaminated segment. The one-dimensional contaminated segment is fed into neural networks. The networks are trained in a supervised, end-to-end manner. The output of neural networks are the cleaned EEG signal. The performance of networks are quantified with multiple metrics as benchmarks.This process was performed separately for ocular artifact removal and myogenic artifact removal.}}
\label{fig:dataset structure}
\end{figure}

\newpage

\newpage

\newpage

\newpage

\newpage

\newpage

\newpage

\newpage
\begin{table}
  \caption{Summary of the data collections used in our dataset}
  \label{table: data summary}
  \centering
  \scalebox{0.7}{
  \begin{tabular}{lllll}
    \toprule

    Dataset   & signal type  & $\#$ of Subjects & mean age $\pm$ SD & dataset website \\
    \midrule
    Cho et al. (2017)  & EEG & $ 52 $& $ 26 \pm 3.86$ & \url{http://gigadb.org/dataset/100295}\\

    Suguru et al. (2016)  & EOG & $20$ & $22.75 \pm 1.45 $ & \url{http://u4ag2kanosr1.blogspot.jp/} \\

    Naeem et al. (2006) & EOG & $8$ & $23.8 \pm 2.5 $ & \url{http://www.bbci.de/competition/iv/} \\
    Schl{\"o}gl et al. (2007) & EOG & $10$ & age between 17 and 31 & \url{http://www.bbci.de/competition/iv/} \\
    Ville et al. (2015) & EMG & $15$ & $40.7 \pm 9.6 $ & \url{http://urn.fi/URN:NBN:fi:tty-201611044685} \\
    \bottomrule
  \end{tabular}}
\end{table}

\newpage
\begin{table}
  \caption{Power ratios of different frequency bands before and after ocular artifact removal}
  \label{table: power of pre/post_EEG}
  \centering
  \begin{tabular}{llllll}
    \toprule
    \cmidrule(r){1-2}
    Denoising method   & delta  & theta  &  alpha & beta & gamma \\
    \midrule
    EMD  & $0.025  $ & $0.042 $ & $\bm{0.096} $ & $0.585 $ & $0.252 $ \\
    Filter & $0.000$ & $0.000 $ & $0.000 $ & $0.405 $ & $0.595 $\\
    
    FCNN  & $0.129  $ & $\bm{0.127} $ & $0.085 $ & $0.500 $ & $\bm{0.159} $ \\
    Simple CNN & $\bm{0.131}$ & $\bm{0.127} $ & $0.085 $ & $\bm{0.492} $ & $0.165 $\\

    Complex CNN & $0.128 $ & $\bm{0.127} $ & $0.085 $ & $0.493 $& $0.166 $  \\
    RNN & $0.124 $ & $0.122 $ & $0.088$  & $0.506 $ & $\bm{0.159}$\\
 
          \hline
    ground truth  & $0.143 $ &  $0.141 $ & $0.093 $ &$0.467 $ &$0.157 $\\
    \hline
    \re{contaminated} signal & $0.514 $ &  $0.216 $ & $0.070 $ &$0.151 $ &$0.049 $ \\
    \bottomrule
  \end{tabular} \\
  \re{\small `Ground truth' refers to the clean EEG segments. \\ `Contaminated signal' refers to the mixed EEG signal generated by Eq. (\ref{Eq: EEG mixing}). \\ \textit{Abbreviation}: EMD, empirical mode decomposition; FCNN, fully-connected neural network; CNN, covolutional neural network; RNN, recurrent neural network.} 
\end{table}

\newpage
\begin{table}
  \caption{Power ratios of different frequency bands before and after myogenic artifact removal}
  \label{table: power of pre/post_EMG}
  \centering
  \begin{tabular}{llllll}
    \toprule
    \cmidrule(r){1-2}
    Denoising method   & delta  & theta  &  alpha & beta & gamma \\
    \midrule
    EMD  & $0.227  $ & $0.162 $ & $\bm{0.093} $ & $0.330 $ & $ 0.188 $ \\
    Filter & $0.000  $ & $0.000 $ & $0.000 $ & $0.312 $ & $0.687 $ \\
    
    FCNN  & $0.147 $ & $0.144 $ & $0.092 $ & $\bm{0.481} $ & $0.135 $ \\
    Simple CNN & $0.119 $ & $0.138 $ & $0.096 $ & $0.506 $ & $0.142 $\\

    Complex CNN & $0.123 $ & $\bm{0.139} $ & $0.097 $ & $0.492 $& $\bm{0.149} $  \\
    RNN & $\bm{0.139} $ & $0.138 $ & $\bm{0.093} $  & $0.482 $ & $0.147 $\\
 
          \hline
    ground truth & $0.142 $ &  $0.140 $ & $0.093 $ &$0.464 $ &$0.160 $\\
    \hline
    \re{contaminated} signal & $0.200 $ &  $0.141 $ & $0.077 $ &$0.300 $ &$0.281 $ \\
    \bottomrule
  \end{tabular}
\end{table}


\end{document}